\documentstyle[prl,aps,psfig]{revtex}
\begin{document}
\twocolumn[\hsize\textwidth\columnwidth\hsize\csname
@twocolumnfalse\endcsname
\title{Stability of money: Phase transitions in an Ising economy}
\author{Stefan Bornholdt and Friedrich Wagner}
\address{Institut f\"ur Theoretische Physik,
Universit\"at Kiel, Leibnizstrasse 15, D-24098 Kiel, Germany}
\maketitle
\begin{abstract}
The stability of money value is an important requisite for 
a functioning economy, yet it critically depends on the 
actions of participants in the market themselves. 
Here we model the value of money as a dynamical variable
that results from trading between agents. The basic trading 
scenario can be recast into an Ising type spin model and
is studied on the hierarchical network structure of a 
Cayley tree. 
We solve this model analytically and observe a phase 
transition between a one state phase, always allowing 
for a stable money value, and a two state phase, 
where an unstable (inflationary) phase occurs. 
The onset of inflation is discontinuous and follows 
a first order phase transition. The stable phase 
provides a parameter region where money value is robust 
and can be stabilized without fine tuning. 
\bigskip
\end{abstract}
]

\section{Introduction}
\label{I}
One of the astonishing facts in economics is the widely 
observed stability of the value of money \cite{Debreu}. 
In models of economic activity often well defined 
mechanisms are introduced that ensure this property, 
e.g.\ by assuming a central agent or market maker 
who supervises the global dynamics and enforces 
market clearance. However, real markets already function 
solely on the basis of the interactions between trading agents 
\cite{Stigler}, raising interesting questions about the validity 
of equilibrium approaches based on central agents \cite{Kirman}. 
The basic dynamics of a decentralized economy can also be studied 
in very simple numerical models of pairwise exchange of goods 
between agents \cite{Donangelo}. Based on pairwise trading alone, 
one observes emergence of money and fluctuations of value without 
any explicit central processes of fixing global variables. 
The value of money appears as a dynamical variable that  
results from the dynamics of trading itself.  

Money as a free parameter in a system of trading agents 
has been studied by Bak, Norrelykke, and Shubik recently \cite{Bak}, 
who cast the problem into a picture consisting of simple  
agents and flows of money and goods between them. 
They place the agents on a line, s.t.\ each trader sells goods to 
his left neighbor and buys products from his right neighbor. 
Combining this system with a periodic boundary condition by 
closing the line to a circle, they observe that the value of money
in general converges to a stable state and emerges as a dynamical 
phenomenon in this setting. They conclude that the general picture 
of this model will also apply to the more complicated heterogeneous 
networks of agents that in general dominate economy.     

However, as the dynamics of this model crucially depend on a 
very specific choice of the boundary condition, 
and as a higher dimensional scenario as well as hierarchies between 
traders may fundamentally change the dynamics, we would like to 
complement this model by a spatial trading model, offering an 
alternative interpretation of Jevons' motivation to understand 
the emergence of money \cite{Jevons}. We will study trading on 
a hierarchical network which allows us to include the interesting 
aspect of hierarchy in the monetary business. Also, moving to higher 
dimensions bears the interesting possibility that a trader with more 
than two neighbors has extra degrees of freedom to optimize himself
by choosing appropriate deals and partners. Finally, we will reformulate
this model in terms of an Ising type spin model that can be solved 
explicitly.

In the next section we will introduce the basic trading model on
a network with dimension greater than one. Section \ref{M} is devoted 
to the problem of competing agents in the presence of a variable 
money value. In section \ref{P} we solve an Ising spin realization 
of the model and study its phase transitions and the conditions 
for a stable value of money.   

\section{A network trading model}
\label{N} 
Let us consider a model where an agent $N$ sells goods which are traded 
via $N-1$ intermediary agents to consumers at level $n=0$. 
This is called the selling mode. The goods are returned by a second chain 
where agent $N$ buys goods, which are traded via $N-1$ different intermediaries from 
$n=0$ (buying mode). Combining both, buying and selling chain, one obtains 
the circular geometry of ref.\ \cite{Bak}. Let us now allow the more 
general scenario that in the selling (or buying) mode each agent can sell 
to (buy from) $z-1$ agents. The linear chains ($z=2$) are replaced by a 
so-called Cayley tree with $z$ neighbors. 
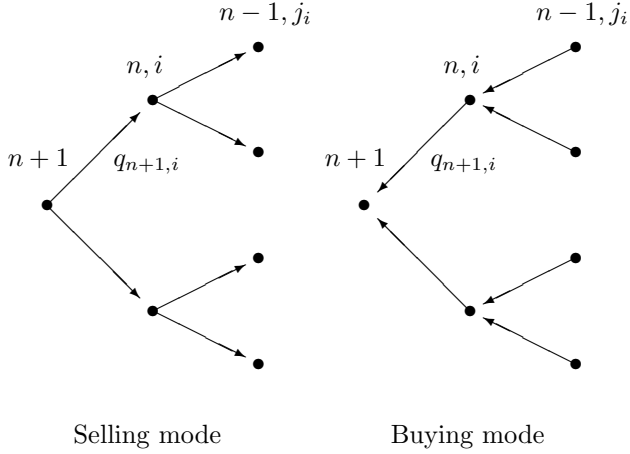
\begin{figure}[htb]
\begin{picture}(220,220)(10,0) 
\put(20,120){\vector(1,1){35}} 
\put(20,120){\vector(1,-1){35}} 
\put(20,120){\circle*{4}} 
\put(60,160){\vector(2,1){35}} 
\put(60,160){\vector(2,-1){35}} 
\put(60,160){\circle*{4}} 
\put(60,80){\vector(2,1){35}} 
\put(60,80){\vector(2,-1){35}} 
\put(60,80){\circle*{4}} 
\put(100,180){\circle*{4}} 
\put(100,140){\circle*{4}} 
\put(100,100){\circle*{4}} 
\put(100,60){\circle*{4}} 
\put(5,135){$n+1$} 
\put(50,170){$n,i$} 
\put(45,135){$q_{n+1,i}$} 
\put(85,190){$n-1,j_i$} 
\put(30,30){Selling mode} 

\put(180,160){\vector(-1,-1){35}} 
\put(180,80){\vector(-1,1){35}} 
\put(140,120){\circle*{4}} 
\put(220,180){\vector(-2,-1){35}} 
\put(220,180){\circle*{4}} 
\put(220,140){\vector(-2,1){35}} 
\put(220,140){\circle*{4}} 
\put(180,160){\circle*{4}} 
\put(220,100){\vector(-2,-1){35}} 
\put(220,100){\circle*{4}} 
\put(220,60){\vector(-2,1){35}} 
\put(220,60){\circle*{4}} 
\put(180,80){\circle*{4}} 
\put(125,135){$n+1$} 
\put(170,170){$n,i$} 
\put(165,135){$q_{n+1,i}$} 
\put(205,190){$n-1,j_i$} 
\put(150,30){Buying mode} 
\end{picture} 
\caption{ Selling mode and buying mode for $z=3$ on a Cayley tree.
The arrows show the flows of goods $q_{n+1,i}$, while money flow 
$g_{n+1,i}$ is opposite to $q_{n+1,i}$ at each link and is not
explicitly shown. }
\label{tree}
\end{figure}
The agents are located at the sites or nodes, while goods and money flow along 
the links of the tree. The agents $(n,i)$ are indexed by the distance 
$n$ from the right hand side of the tree. The index $i$ distinguishes 
different agents at the same distance and will be written only if necessary. 
For the amount of goods $q_{n,i}$ flowing between agents 
$n$ and $n-1,i$ we use the normalized variable 
\begin{eqnarray}
\bar{q}_{n,i} = \frac{1}{(z-1)^{n-1}} \; q_{n,i}
\label{normalized}  
\end{eqnarray}
The amount of traded goods is described by two utility functions. 
If an agent $n$ sells ${q}$ at the price $p$ he gains the utility 
\begin{eqnarray}
u_n^{(S)} = I_n \; \bar{q} \; p - \tilde{c}(\bar{q}). 
\label{utilitys}  
\end{eqnarray}
$I_n$ denotes the value of money and $\tilde{c}(\bar{q})$ 
the decrease of $u$ 
by losing $q$. Similarly, if the agent buys ${q}$ at price $p$ 
the utility reads   
\begin{eqnarray}
u_n^{(B)} = d(\bar{q}) - I_n \; \bar{q} \; p \; . 
\label{utilityb}  
\end{eqnarray}
It is important to use the normalized flow of goods (\ref{normalized})  
instead of q for the following reason. In the monopolistic equilibrium 
all the money values $I_n$ are the same and all goods are conserved. 
Therefore, the goods $q_n$ increase with $(z-1)^{n-1}$. 
Then, the utilities (\ref{utilitys}) and (\ref{utilityb}) express the 
assumption that an agent level of $n$ gets the same utility by trading 
$q_n=(z-1) \; q_{n-1}$ as the agents at level $n-1$ trading $q_{n-1}$. 
As in ref.\cite{Bak} each agent $n$ can choose its own money value 
$I_n$, the amount of bought goods ${q}_{n+1}$, and the price $p_{n,i}$ 
for sold goods ${q}_{n,i}$. 
For a meaningful problem one has to use the following assumption, 
also made in ref.\ \cite{Bak}: The time scale on which the $q$ 
or $p$ change is much shorter than the scale of changing the money values. 
Therefore we can optimize the coupled system (\ref{utilitys}) and 
(\ref{utilityb}) with fixed values of $I_n$. An additional dynamics 
must be used for finding $I_n$ from $q(I)$ and $p(I)$. For the utility 
functions ${d}$ and $\tilde{c}$ power laws have been used in ref.\
\cite{Bak}. This property  is not really needed. It is sufficient for 
$\tilde{c}$ to increase faster than, and for ${d}$ less than linearly 
for large $\bar{q}$. To avoid algebraic complications, we here use for  
${d}$ a power law 
\begin{eqnarray}
{d}(x) = \frac{1}{\beta} x^\beta \;\;\;\mbox{with}\;\;\; \beta<1  
\label{dtilde}  
\end{eqnarray}
and for $\tilde{c}$ a power law only in the example of section \ref{P}
\begin{eqnarray}
\tilde{c}(x) = \frac{1}{\alpha} x^\alpha \;\;\;\mbox{with}\;\;\; \alpha>1.  
\label{ctilde}  
\end{eqnarray}
In the general case, $\tilde{c}$ must have positive first and second 
derivative. Having performed the optimization all quantities can be
expressed by the Legendre transform of $\tilde{c}(s^{1/\beta})$ 
denoted by $c(r)$. For the power law (\ref{ctilde}) we get
\begin{eqnarray}
c(r)= \frac{\alpha-\beta}{\alpha \cdot \beta}
(\beta r)^\frac{\alpha}{\alpha-\beta}.
\end{eqnarray}
The optimization is slightly different in the buying or selling mode. 
In the latter we have for each agent $n$ 
\begin{eqnarray}
u_n^{(B)} &=& \frac{1}{\beta}\; \bar{q}_{n+1}^\beta - I_n \; 
\bar{q}_{n+1} \; p_{n+1} \;\;\;\;\;\; n=0,\dots,N-1 
\label{utilityb2}  
\\ 
u_n^{(S)} &=& I_n \; \sum_i \; \left[ \bar{q}_{ni} \; p_{ni} 
- \tilde{c}(\bar{q}_{ni}) \right] \;\;\;\;\;\; n=1,\dots,N. 
\label{utilitys2}  
\end{eqnarray}
Since $\sum_i \tilde{c}(\bar{q}_i) < \tilde{c}(\sum_i \bar{q}_i)$ 
the agents $n$ will handle each selling to agents $n-1,i$ separately, 
and not lump all requests $q_{ni}$ into a single order. The optimization
begins at $n=0$, where only $u_n^{(B)}$ is present. The maximum of 
(\ref{utilityb2}) leads to the value $\bar{q}_1$. This value
$\bar{q}_{1i}(p_{1i})$ is used to optimize $p_{1i}$ in $u_1^{(S)}$ given by
(\ref{utilitys2}). This procedure is repeated to the top agent $N$.
The resulting values of traded goods and money flow $g_{n,i}$
from $n-1,i$ to $n$
\begin{eqnarray}
g_{n,i} = q_{n,i} \; p_{n,i} 
\end{eqnarray}
are given by 
\begin{eqnarray}
q_{n,i}&=& (z-1)^{n-1} \; 
\left[ c' \left( \frac{I_n}{I_{n-1,i}} \right) \right]^\frac{1}{\beta} 
\\ 
g_{n,i}  &=& (z-1)^{n-1} \; \frac{1}{I_{n-1,i}} \;  
\; c' \left( \frac{I_n}{I_{n-1,i}} \right).  
\end{eqnarray}
One sees that the goods flow and the valued money flow 
$I_{n,i}\;g_{n,i}$ only depend on the ratios
$I_n/I_{n-1}$, but not on the absolute scale of $I$. 
The value of utilities in (\ref{utilityb2}) and 
(\ref{utilitys2}) at the maximum are given by  
\begin{eqnarray}
u_n^{(B)} = \frac{1-\beta}{\beta} \;\; c'\left(\frac{I_{n+1}}{I_n}\right) 
\label{utilityb3}  
\end{eqnarray}
and by
\begin{eqnarray}
u_n^{(S)} = \sum_{i=1}^{z-1} \;\; c\left(\frac{I_n}{I_{n-1,i}}\right). 
\label{utilitys3}  
\end{eqnarray}
The buying mode can be treated with the same method. 
It can be obtained from the selling mode by interchanging 
at each link $(n;n-1,i)$ the adjacent $I_n$ and $I_{n-1}$. This leads to 
\begin{eqnarray}
q_{n,i}= (z-1)^{n-1} \; 
\left[ c' \left( \frac{I_{n-1,i}}{I_n} \right) \right]^\frac{1}{\beta} 
\nonumber \\ 
g_{n,i}= (z-1)^{n-1} \; \frac{1}{I_n} \;  
 c' \left( \frac{I_n-1,i}{I_n} \right)  
\end{eqnarray}
and  the utility functions at maximum 
\begin{eqnarray}
u_n^{(B)} &=& \frac{1-\beta}{\beta} \; \sum_{i=1}^{z-1} \; 
c'\left(\frac{I_{n-1,i}}{I_n}\right) 
\nonumber \\ 
u_n^{(S)} &=& c\left(\frac{I_n}{I_{n+1}}\right). 
\label{utilit4}  
\end{eqnarray}
In the case of the linear chain ($z=2$) the only 
difference between selling and buying is a reordering of $I$, 
which is performed in \cite{Bak} by placing the agents on a 
circle. We can consider the normalized money ratio   at site 
$n$ given by (in the selling mode) 
\begin{eqnarray}
\Delta g(n+1,n,i) &=& (z-1)\frac{ g_{n-1,i}}{g_{n}} 
\nonumber \\ 
&=&   \frac{I_n}{I_{n-1,i}} \;\;\left[ c'\left( \frac{I_n}{I_{n-1,i}} 
\right) \Big / c' \left( \frac{I_{n+1}}{I_n} \right) \right]. 
\label{balance}  
\end{eqnarray}
Note that $\Delta g$ is only a function of the ratios
\begin{eqnarray}
r_{n,i} = \frac{I_{n+1}}{I_{n,i}}.
\label{ratio}
\end{eqnarray}
Further, money conservation at agent $n$ implies
\begin{eqnarray}
\sum_{i=1}^{z-1}\Delta g(n+1,n,i)=z-1. 
\label{conse}
\end{eqnarray}
For $r>1$ money is accumulated at agent $n$, while $r<1$ 
means that money has to be borrowed. Therefore in the selling mode $r<1$ 
implies an inflation, while values $r>1$ imply deflation. 
In the buying mode $r$ is essentially replaced by $1/r$ such that 
$\Delta g$ is 
given by 
\begin{eqnarray}
\Delta g(n+1,n,i) &=&  \frac{I_{n+1}}{I_{n}} \;\; \left[
 c'\left( \frac{I_{n-1,i}}{I_{n}}
\right)\Big  / c' \left( \frac{I_{n}}{I_{n+1}} \right) \right]
\label{balancebuy}
\end{eqnarray}
and the reversed statements are true. 

In the case of $z=2$ in \cite{Bak} the condition of
money conservation has been applied. Both, the strategy of  
storing money (`Dagobert Duck mode'), as well as the strategy of 
spending unlimited amounts of money (`Donald Duck mode') are punished. 
There in each step of the update of $I_n$ the condition 
\begin{eqnarray}
\Delta g \;=\; 1
\label{equilibrium}  
\end{eqnarray}
is imposed. The change of $I_n$ results in new $q,p$ values and 
the procedure is repeated until convergence to (\ref{equilibrium}) 
is reached (the additional delay in \cite{Bak} only changes the 
time scale, but not the equilibrium (\ref{equilibrium})).   

For $z>2$ money conservation involves a sum of $\Delta g$ over $i$. 
To fix the money flow $g_{n,i}$ to agent $n$, extra conditions are needed. 
Such a condition may result from the cooperation between agents
$n-1,i$ connected to $n$. Suppose agent $n$ sells the amount 
$q=\sum_iq_{n,i}$ 
which is bought by agents $n-1,i$. If they do not cooperate, one agent 
may choose its $I_{n-1,i}$ such that the sum is exhausted. Then the system 
will collapse into a linear chain. If they cooperate, 
they optimize their common utility 
\begin{eqnarray}
u^{(B)} = \sum_i \; \frac{1-\beta}{\beta} \; \bar{q}_{n,i}^\beta 
\label{utilityb4}  
\end{eqnarray}
as function of $I_{n-1,i}$ subject to the condition 
$\sum q_{n,i} =q$ since the 
$q_{n,i}$ are unique functions of the $I_{n-1,i}$.  
For $\beta<1$ $u^{(B)}$ has a maximum for equal $q_{n,i}$ 
which implies $I_{n-1,i}$ is independent of $i$. 
Therefore we have the condition (\ref{equilibrium}) 
also valid for $z>2$. 
In terms of the ratios $r$ it reads 
\begin{eqnarray}
c'\left( r_{n} \right) = r_{n-1} \; \; c'\left( r_{n-1} \right). 
\label{cost}  
\end{eqnarray}
This  recursion formula for the ratios $r_n$
exhibits the stable fixed point $r_n=1$, since both $c'(1)$ and $c''(1)$
are positive.
The value $r_{N-1}$ is arbitrary. After a transient region, the $r_n$ 
for $n\ll N$ are equal $1$. For power laws the recursion can be solved 
explicitly. $r_n$ depends only on the ratio of the exponents
\begin{eqnarray}
\gamma =\frac{\beta}{\alpha} 
\end{eqnarray}
which can be called the relative elasticity of the utility functions, 
and is given by
\begin{eqnarray} \label{rsell}
\log r_n=\gamma^{N-1-n}\log r_{N-1}\;.
\end{eqnarray}
The same method can be applied in the buying mode. Now $r'_0$ can be chosen
arbitrarily due to the replacement $r_n \rightarrow 1/r_n$
\begin{eqnarray} \label{rbuy}
\log r_n'=\gamma^{n}\log r_{0}'\;.
\end{eqnarray}
Both (\ref{rsell}) and (\ref{rbuy}) can be used to obtain $I_n$ resp. $I_n'$
for the buying mode. In the  selling mode, money is 
accumulated at agent $N$
and the agents at $n=0$ have to borrow money. In order to
`recycle' the money, one can connect $n=0$ and $n=N$ with a
second tree in the buying mode where agents $n=0$ sell other goods
$q'$ over this second tree to agent $N$. From $I_0=I_0'$ and $I_N=I_N'$
the constants $r_{N-1}$ and $r'_0$ can be eliminated with the result
\begin{eqnarray}
I_n = I_0 \; \left( \frac{I_N}{I_0} \right)^{\gamma^{n-N}}.
\label{solution}
\end{eqnarray}
\begin{eqnarray}
I'_n = I_N \; \left( \frac{I_0}{I_N} \right)^{\gamma^n}.
\label{solution2}
\end{eqnarray}
In both (\ref{rsell}) and (\ref{rbuy}) terms $\gamma^N \ll 1$ have been
neglected in the exponent.
\begin{figure}[htb]
\let\picnaturalsize=N
\def\picsize{90mm}
\def\picfilename{fig2.ps}
\ifx\nopictures Y\else{\ifx\epsfloaded Y\else\input epsf \fi
\let\epsfloaded=Y
\centerline{\ifx\picnaturalsize N\epsfxsize \picsize\fi
\epsfbox{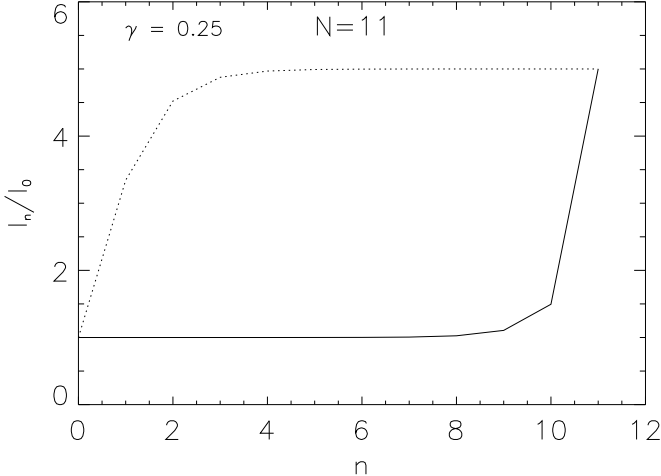}}}\fi
\medskip
\quad 
\caption{The money values $I_n/I_0$ for the selling mode (solid line)
         and $I_n/I_0$ for the buying mode (dotted line) as function
         of $n$. A seller dominated market with $I_N/I_0=5$ has been
         assumed. The increase of $I_n/I_0$ near $N$ exhibits the
         ``peanuts effect''. }
\label{peanuts}
\end{figure}
The money values $I_N$ of agent $N$ and $I_0$ of the agents at $n=0$ are free
constants. Their choice depends on the relative weight the
agents place on the utilities in the buying or selling mode.
A seller dominated market leads to $I_N > I_0$.
In Fig.\ \ref{peanuts} we show $I_n$ and $I_n'$ as function
of $n$ for $\gamma=1/4$ and $N=11$. $I_n$ ($I_n'$) are constant
over a wide range and change in the last (first) two steps to
the values imposed by the boundary conditions. Constant money values
are achieved even when they are different in the selling and buying mode. 
This shows that the assumption of periodic boundary conditions made in
ref.\ \cite{Bak} is crucial for constant money values derived
from money conservation $\Delta g =1$ and not just minimizing the
finite size effects as in physical problems.

Another consequence of the recursion is the ``peanuts effect'':
Consider the normalized flows of goods
$\bar{q}_n$ in a seller dominated market, using
a power law ansatz for utility functions. 
They are constant for $n \ll 1$ and increase near $n=N$. The ratio
\begin{eqnarray}
\frac{\bar{q}_N}{\bar{q}_0} = \left(\frac{I_N}{I_0}\right)^{1/(\alpha -\beta)}
\end{eqnarray}
may take large values, such that also $u_N^{(S)}/u_0^{(B)}$ becomes
large. This remarkable feature seems to have induced an unfortunate 
german banker to publicly call the credits given to small customers 
at $n=0$ as ``peanuts'' (a statement that was not agreed upon by the 
broad public) \cite{Peanuts}.

Up to now, the number $z-1$ of neighbors $n-1$ adjacent to agent $n$
did not play any r\^{o}le given their money value $I_{n-1,i}$
has been chosen equally. In the next chapter we use a dynamics
to reach the equilibrium condition (\ref{equilibrium}) from an
arbitrary initial state, including thermal noise.
The utility function for updating the $I_n$ may have other maxima
besides the maximum described by (\ref{equilibrium}).
This we investigate in the next section.

\section{Utility function for money values}
\label{M} 
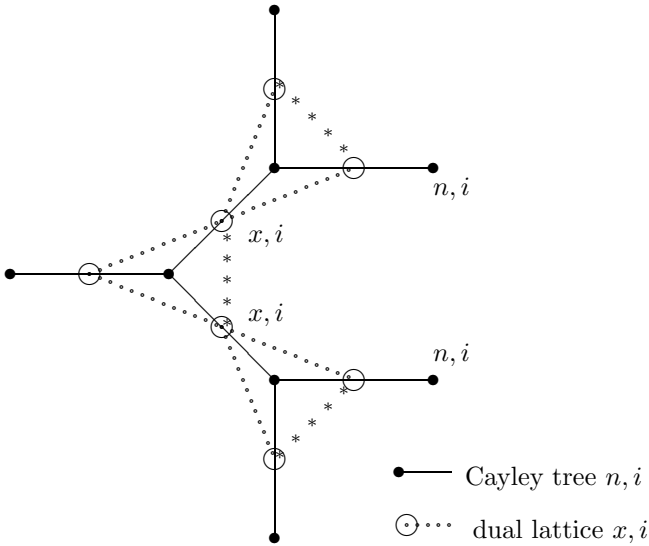
\begin{figure}[htb]
\begin{picture}(220,220)(10,0) 

\put(0,120){\line(1,0){60}} 
\put(60,120){\line(1,1){40}} 
\put(60,120){\line(1,-1){40}} 
\put(100,160){\line(0,1){60}} 
\put(100,160){\line(1,0){60}} 
\put(100,80){\line(1,0){60}} 
\put(100,80){\line(0,-1){60}} 

\put(0,120){\circle*{4}} 
\put(60,120){\circle*{4}} 
\put(100,20){\circle*{4}} 
\put(100,80){\circle*{4}} 
\put(100,160){\circle*{4}} 
\put(100,220){\circle*{4}} 
\put(160,160){\circle*{4}} 
\put(160,80){\circle*{4}} 

\put(30,120){\circle{8}} 
\put(80,140){\circle{8}} 
\put(80,100){\circle{8}} 
\put(100,50){\circle{8}} 
\put(100,190){\circle{8}} 
\put(130,80){\circle{8}} 
\put(130,160){\circle{8}} 

\multiput(30,120)(4,1.6){13}{\circle{1}} 
\multiput(30,120)(4,-1.6){13}{\circle{1}} 
\multiput(80,100)(0,8){5}{\scriptsize $\ast$ \normalsize}

\multiput(80,140)(1.6,4){13}{\circle{1}} 
\multiput(80,140)(4,1.6){13}{\circle{1}} 
\multiput(100,190)(6.05,-6.05){5}{\scriptsize $\ast$ \normalsize}

\multiput(80,100)(1.6,-4){13}{\circle{1}} 
\multiput(80,100)(4,-1.6){13}{\circle{1}} 
\multiput(100,50)(6.05,6.05){5}{\scriptsize $\ast$ \normalsize}

\put(172,40){Cayley tree $n,i$} 
\put(147,45){\circle*{4}} 
\put(147,45){\line(1,0){20}} 

\put(175,20){dual lattice $x,i$} 
\put(150,25){\circle{8}} 
\multiput(150,25)(4,0){5}{\circle{1}} 

\put(90,132){$x,i$}
\put(90,102){$x,i$}
\put(160,150){$n,i$}
\put(160, 87){$n,i$}

\end{picture}
\caption{The Cayley tree and its dual lattice for $z=3$. On the links
denoted by * the $\Delta g$ variables are absent.}
\label{dual}
\end{figure}

The dynamics of ref.\cite{Bak} for the money value $I_n$ 
is based on the conservation
of money flux expressed by $\Delta g =1$ in the case $z=2$.
This method has several disadvantages. It is completely deterministic
and does not allow for noise. More importantly, it does not
involve the agents whose utility functions are minimal
for the monopolistic equilibrium $r=1$. Even a possible
utility function for the dynamics would be rather complicated,
since $\Delta g$ on a Cayley tree connects agent $n+1$ with agents $n-1,i$
corresponding to a next to next neighbor interaction.
In addition, we encounter for $z>2$ the difficulty that money 
conservation in eqn.\ (\ref{conse}) does not determine the dependence
of $g_{n,i}$ on $i$. To improve and to generalize the method of \cite{Bak}
the dynamics of the money values will be based on an
utility function $H$. Then the noise can be described by a
Boltzmann distribution.
$H$ is the sum of two parts: One part $H_M$ contains the effect
of the money authorities, the second $H_A$ is due to the agents.
The latter should involve all agents equally. The simplest choice
corresponds to a sum over all utilities $u^{S}+u^{B}$.
The key observation is that the utilities depend on variables $q_{n,i}$ or 
$r_{n-1,i} = I_n/I_{n-1,i}$, which are defined on the links
$x=(n; n-1,i)$ of the lattice. Moreover, the sum of utilities
can be rearranged into a sum over links $x$
\begin{eqnarray}
H_A &=& \sum_{agents} \; u^{S}+u^{B} \cdot \nonumber \\
    &=&  \sum_{ x} \; u_A(r_{x})
\end{eqnarray}
with $u_A$ given in the selling mode by 
\begin{eqnarray}
u_A(z) &=& c(z) + \frac{1-\beta}{\beta} \; c'(z).
\end{eqnarray} 
The money authority part must favor $\Delta g(n+1,n,i)=1$.
This establishes money conservation and a certain cooperation
of the agents $n,i$ to prefer equal money values $I_{n,i}$.
Since due to equation (\ref{balance}) $\Delta g=1$
only involves neighboring ratios, this suggests that one should 
consider the model on the dual lattice
which is obtained by replacing the links $(n+1; n)$ 
of the Cayley tree   
by nodes ${x}$, and the nodes $n$ by $z-1$ dimensional 
hypertetraeders. This dual lattice for $z=3$ is called a cactus
and is depicted in Fig.\ \ref{dual}. 
\noindent
$\Delta g(x,y)$ are variables defined on the links $x,y$ of the dual
lattice. Nonvanishing values of $\Delta g$ exist only on the links $x,y$ 
depicted by the dotted lines in Fig.\ \ref{dual} which are
denoted by $x>y$. We model $H_M$ by a sum over all links $x>y$
of a utility function $u_M(\Delta g)$ having a maximum at
$\Delta g =1$. So we arrive at the following utility function for 
the dynamics of $r$:
\begin{eqnarray}
H(r) &=& \sum_{x}
u_A(r_{x}) \; +\;  \sum_{x>y} \; u_M(\Delta g(x,y)). 
\label{h}
\end{eqnarray}
Possible equilibrium states without noise correspond to the 
maximum of (\ref{h}). Thermal noise is introduced by 
assuming that the equilibrium distribution $w(r)$ for $r$ is Boltzmann
distributed with the utility function (\ref{h})
\begin{eqnarray}
w(r) \sim e^{\beta_T \; H(r)}. 
\label{dist}
\end{eqnarray}
$\beta_T$ corresponds to the inverse temperature and 
$\beta_T\rightarrow\infty$ would be the deterministic limit. 
There exist many dynamics having (\ref{dist}) as equilibrium 
distribution. Particular interesting are local algorithms as the
Glauber  \cite{Glauber} dynamics
or the Metropolis algorithm \cite{Metropolis}. 
In the latter a randomly chosen agent $n$ 
selects a new $I'_n$ thereby changing its neighboring ratios 
$r_x$ to $r'_x$. The change $I'_n$ is accepted with probability 
\begin{eqnarray}
p= e^{\beta_T \; \mbox{min}(0, \Delta H)}   
\end{eqnarray}
where $\Delta H = H(r') - H(r)$ denotes the change in the utility 
function (\ref{h}). It only involves the neighboring $r_x$ which are known 
to the agents by the money- and goods flows at $n$. 

In the following section we discuss a realization of (\ref{h})
within an Ising type model. There has been a tradition of using 
Ising and similar spin models in economic theory \cite{Ising}. 
Here, using an Ising formulation has the advantage that the 
probabilities (\ref{dist}) for the average $r_x$ or correlations
can be computed explicitly.

\section{Phase transitions in the Ising model}  
\label{P} 
\begin{figure}[htb]
\let\picnaturalsize=N
\def\picsize{90mm}
\def\picfilename{fig4.ps}
\ifx\nopictures Y\else{\ifx\epsfloaded Y\else\input epsf \fi
\let\epsfloaded=Y
\centerline{\ifx\picnaturalsize N\epsfxsize \picsize\fi
\epsfbox{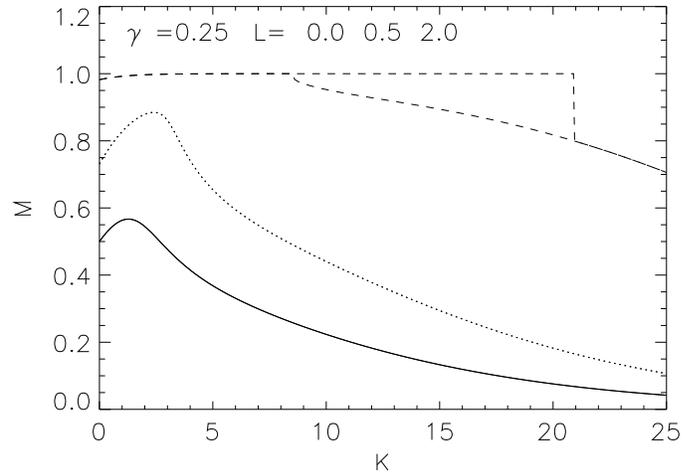}}}\fi
\quad 
\caption{The inflation parameter $M$ as a function of $K$ for 
         $\gamma=0.25$ and $z=3$. The $L$ values are 0 (solid line), 
         0.5 (dotted line), and 2.0 (dashed line). For $L=0,0.5$ 
         the system is in the one state phase. For $L=2$ and 
         $8.5<K<20.9$ the system allows two possible fixed 
         points (two state phase). }
\label{magnetization_a}
\end{figure}

\begin{figure}[htb]
\let\picnaturalsize=N
\def\picsize{90mm}
\def\picfilename{fig6.ps}
\ifx\nopictures Y\else{\ifx\epsfloaded Y\else\input epsf \fi
\let\epsfloaded=Y
\centerline{\ifx\picnaturalsize N\epsfxsize \picsize\fi
\epsfbox{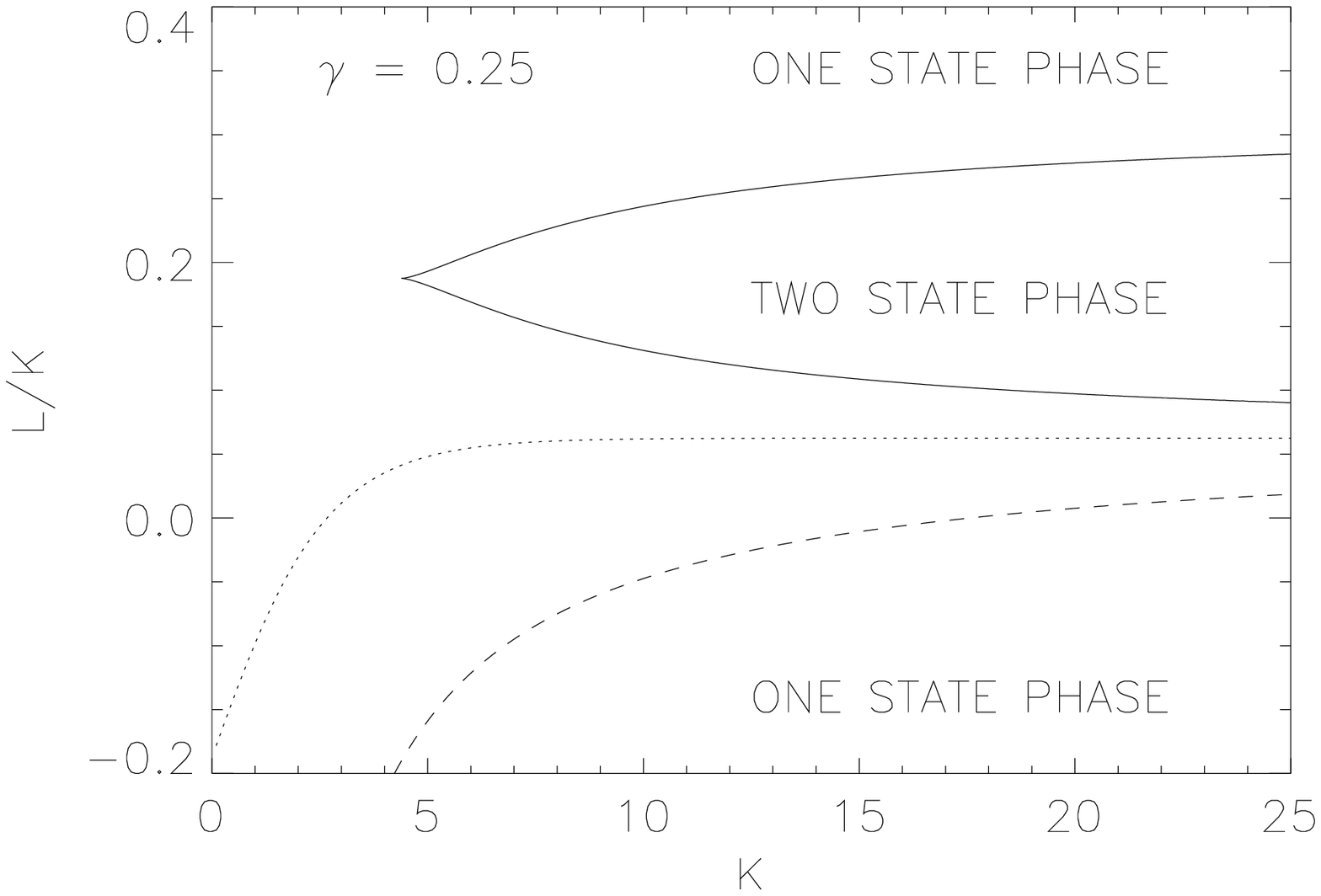}}}\fi
\quad 
\caption{Phase diagram for $\gamma=0.25$ and $z=3$ in the
         $(L/K,K)$ plane. The solid lines show the critical curves
         $L_\pm(K)$ where the system changes from the OSP to the TSP
         $(L_-(K) < L < L_+(K)$. Along the dotted (dashed) line 
         $M=0.5$(0.1) holds. Above the dotted (dashed) line 
         $M>0.5$(0.1), below $M<0.5$(0.1). }
\label{phase_a}
\end{figure}

\begin{figure}[htb]
\let\picnaturalsize=N
\def\picsize{90mm}
\def\picfilename{fig7.ps}
\ifx\nopictures Y\else{\ifx\epsfloaded Y\else\input epsf \fi
\let\epsfloaded=Y
\centerline{\ifx\picnaturalsize N\epsfxsize \picsize\fi
\epsfbox{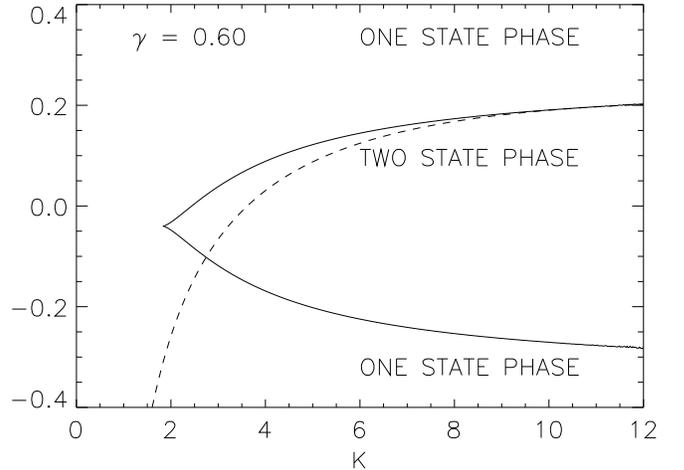}}}\fi
\quad 
\caption{Phase diagram for $\gamma=0.6$ and $z=3$ in the
         $(K,L/K)$ plane. The solid lines show the critical curves
         $L_\pm(K)/K$ as in Fig.\  \ref{phase_a}. Along the
         the dashed line one solution for $w$ leads to $M=0.1$.
         Below this line $M$ can be smaller than 0.1.}
\label{phase_b}
\end{figure}

In the deterministic limit $\beta_T\rightarrow \infty$ the utility 
function (\ref{h}) should lead to the state $r_{{x}}=1$, 
corresponding to the absolute maximum of $H$. However, there 
may be additional local maxima with $r\not=1$ which are frozen 
if the thermal noise vanishes.
To study this possibility we consider the following
simplified version of (\ref{h}). We allow only small deviations 
of $r$ from $1$ and parametrize $r$ by a two valued function 
with one value 1 and the other $r_0$ close to 1
\begin{eqnarray}
r_{{x}} = r_{{0}}^{\frac{1+\sigma_{{x}}}{2}}
\label{twovalued}
\end{eqnarray}
with an Ising spin variable $\sigma_x=\pm 1$. In addition 
we assume for the utility function $c$ a power law as 
in (\ref{ctilde}). The Boltzmann weight (\ref{dist}) 
is a product of site factors 
\begin{eqnarray}
G^{(0)}(\sigma_{{x}}) = 
e^{\beta_T \; u_A(\sigma_x)} 
\end{eqnarray}
and link factors 
\begin{eqnarray}
G^{(1)}(\sigma_{{x}},\sigma_{{y}}) = 
e^{\beta_T \; u_M(\Delta g)}  
\end{eqnarray}
with $\Delta g$ derived from (\ref{balancebuy}) for the buying mode
and from (\ref{balance}) for the selling mode. In the latter we obtain
\begin{eqnarray} 
\Delta g (\sigma_{{x}},\sigma_{{y}}) = 
r_0^{(1-\gamma + \sigma_{y} - \gamma \sigma_{x})/(2(1-\gamma ))}. 
\end{eqnarray}
For $r_0$ close to $1$, we can expand $u_M$ around $1$ and obtain 
\begin{eqnarray}
G^{(1)}(\sigma_{{x}},\sigma_{{y}}) = 
\left( \begin{array}{cc}
e^{-K(1-\gamma)^2} & e^{-K\gamma^2} \\ e^{-K} & 1 
\end{array} \right)_{\sigma_{{x}},\sigma_{{y}}} 
\label{matrix}
\end{eqnarray}
with the money conservation constant 
\begin{eqnarray}
K = \left( -\frac{\beta_T \; u''_M(1)}{2} \right) \; 
\left(\frac{\alpha \ln r_0}{\alpha-\beta} \right)^2. 
\end{eqnarray}
In (\ref{matrix}) the irrelevant factor $\exp(\beta u_M(1))$ 
has been omitted. In the same way we obtain for $G^{(0)}$ 
\begin{eqnarray}
G^{(0)}(\sigma_{{x}}) = 
e^{(z-1) \; L \; \delta_{\sigma_{{x}},1}}  
\label{g0} 
\end{eqnarray}
with the self interest constant 
\begin{eqnarray}
L = \frac{\beta_T }{z-1}  \; \left[ u_A(r_0)-u_A(1) \right].
\end{eqnarray}
Using (\ref{matrix}) and (\ref{g0}) the Boltzmann equilibrium distribution 
for the dynamical variables $\sigma$ can be written as 
\begin{eqnarray}
w(\sigma) = \frac{1}{Z} \; \prod_{{x}} \; G^{(0)}(\sigma_{{x}}) 
\; \prod_{{y}<{x}} \; G^{(1)}(\sigma_{{x}},\sigma_{{y}}). 
\end{eqnarray}
The normalization factor $Z$ follows from the condition 
$\sum_{\{\sigma\}}w(\sigma)=1$. The distribution for a single spin 
$w_1(\sigma_x) = \sum_{\sigma\not= \sigma_x}  w(\sigma)$ or two spins 
$w_2(\sigma_x\sigma_y) = \sum_{\{\sigma\not= \sigma_x,\sigma_y\}}  w(\sigma)$ 
can be calculated recursively \cite{Wagner}. 
For this purpose we introduce the tree distribution 
$T_n(\sigma_x)$ of length $|x| = n$ corresponding to the product 
of all factors $G^{(0)}$ and $G^{(1)}$ on a dual tree starting at 
${x}$, which is summed over all spins $\sigma_{{y}}$ with $|y|\ge |x|$
\begin{eqnarray}
T_n(\sigma_{{x}}) &=&\frac{1}{Z_T}\;\sum_{\{\sigma_{y},\;{y}>{x}\}} 
\; \prod_{|y|\geq |x|}\cdot  
\nonumber \\ 
&& \left( G^{(0)}(\sigma_{{y}})\;
\prod_{|y'| \geq |y|} \; G^{(1)}(\sigma_{{y}},\sigma_{{y}'})\right). 
\label{Tn} 
\end{eqnarray}
$Z_T$ is chosen such that $\sum_\sigma T(\sigma)=1$. 
For agent $N$ (\ref{Tn}) yields the equilibrium distribution 
$w_1(\sigma_N)$, 
for agents $|x| < N$ the tree distribution $T_n(\sigma_x)$ is a 
conditional probability related to $w_1(\sigma_{{x}})$.
According to (\ref{Tn}) a tree of length $n$ can be expressed by 
trees of length $n-1$ in the following way 
\begin{eqnarray}
T_n(\sigma)=G^{(0)}(\sigma) \; \sum_{\sigma_1, \dots \sigma_{z-1}} 
\; \prod_{i=1}^{z-1} \; G^{(1)}(\sigma,\sigma_i) \; T_{n-1}(\sigma_i). 
\label{Tn2} 
\end{eqnarray}
Any function $T(\sigma)$ depending on a variable 
$\sigma = \pm 1$ can be parametrized as 
\begin{eqnarray}
T_n(\sigma)= a_n\; \left(w_n \; \delta_{\sigma,1} + \delta_{\sigma,-1} \right). 
\label{Tn3} 
\end{eqnarray}
Carrying out the summation in (\ref{Tn2}) we find a recursion relation 
for $a_{n+1}$ and $w_{n+1}$ in terms of $a_n$ and $w_n$. For the latter
this reads  
\begin{eqnarray}
w_{n+1} &=& f(w_n) \\ f(w) &=& 
\left[ e^{-K\gamma^2+L} \;\; \frac{1+e^{(2\gamma-1)K}\;w}{1+e^{-K}\;w} 
\right]^{z-1}
\label{wrecursion} 
\end{eqnarray}
which allows the recursive calculation of $w_n$ if the values $w_n$ 
at $n=0$ are given. The mean value of $r_x$ for the top agent $x=N$ 
is related to $w_N$ by the inflation parameter  
\begin{eqnarray}
M= \left\langle \frac{\ln r_x}{\ln r_0} \right\rangle  
= {\langle \delta_{\sigma_x,1} \rangle}_T  
= \frac{w_N}{1+w_N}.   
\label{mtree} 
\end{eqnarray}
Therefore $w_N\sim 0$ expresses the preference for $r_x=1$, 
whereas $w_N \gg 1$ leads to $r_x\sim r_0$. In physical
problems $2M-1$ corresponds to the magnetization. Since our
utility function is not symmetric under $\sigma \rightarrow -\sigma$
the disordered state $M=1/2$ has no particular meaning.
Here only the fully magnetized states are interesting.
Inflation parameter $M=0$ corresponds to the monopolistic state
and $M=1$ implies inflation ($r_0<1$) or deflation ($r_0>1$).
Particularly interesting are stable fixed points $w_n$ of 
the recursion (\ref{wrecursion}). These are solutions independent 
of $n$ for $n \gg 1$, especially independent of the boundary 
values $w_0$. They correspond to a homogeneous value of the 
inflation parameter on the lattice. If more than one fixed point
exists, the system can exhibit different phases. It is a particular
property of the Cayley tree   that the values at the boundary
decide which phase is adopted \cite{Recursion,Wagner}. On a normal
finite dimensional lattice only one phase would be thermodynamically
stable. The form of $f(w)$ shows that the fixed point equation 
$w=f(w)$ can have either one or two solutions satisfying the stability
condition $|f'(w)|<1$. Depending on the values of $K,L$ and
$\gamma$ there can be a one state phase (OSP) with a unique
value of $M$ or a two state phase (TSP) with two possible
values. In Fig.\ \ref{magnetization_a} 
we show  for the numerical solution of $w=f(w)$ with $z=3$ neighboring
agents the inflation parameter 
$M(w)$ as a function of $K$ for several $L$ values and $\gamma=0.25$. 
\noindent
At low $L$ there will be a unique solution (OSP) in which $M$ tends
to zero for large values $K$ of the money conservation term in $H$.
$M$ increases with the self interest $L$ of the agents.
For sufficiently large $L$ a switch into the TSP with two 
possible values of $M$ occurs. Still the monopolistic equilibrium
can be achieved for large $K$. The fixed point equation
can be only solved numerically, the calculation of the phase boundaries 
requires solution of quadratic equations. One finds that
TSP only occurs if the following two conditions are satisfied.
$K$ has to be larger than a critical value given by
\begin{eqnarray}
K_c = \frac{1}{\gamma} \; \ln \frac{z}{z-2} \; ,
\label{Kcrit}
\end{eqnarray}
and $L$ has to be bounded by
\begin{eqnarray}
L_-(K) < L < L_+(K) \; .
\label{bounds}
\end{eqnarray}
For the following we need only the asymptotic form of $L_{\pm}(K)$
for $K \gg 1$
\begin{eqnarray}
L_\pm = K \; \left( \gamma^2+\frac{1}{z-1} \right) - 2\gamma K \;
\left\{ \begin{array}{l} \frac{1}{z-1} \\ 1 \end{array} \right.
\end{eqnarray}
For the linear chain ($z=2$) discussed in \cite{Bak}
$K_c$ becomes infinite and a phase transition to TSP
cannot occur. The second condition (\ref{bounds}) explains
that a window for the TSP phase is observed in 
Fig.\ \ref{magnetization_a}. Another feature of the model is
the dependence on the elasticity $\gamma$. 
For values of $\gamma < \gamma_c$
with
\begin{eqnarray}
\gamma_c = 1-\sqrt{\frac{z-2}{z-1}}
\end{eqnarray}
the lower bound $L_-$ is always positive. For fixed $L$ and
$K\rightarrow\infty$ one always ends up in the OSP in
agreement with what we have seen in Fig.\ \ref{magnetization_a}.
\begin{figure}[htb]
\let\picnaturalsize=N
\def\picsize{90mm}
\def\picfilename{fig5.ps}
\ifx\nopictures Y\else{\ifx\epsfloaded Y\else\input epsf \fi
\let\epsfloaded=Y
\centerline{\ifx\picnaturalsize N\epsfxsize \picsize\fi
\epsfbox{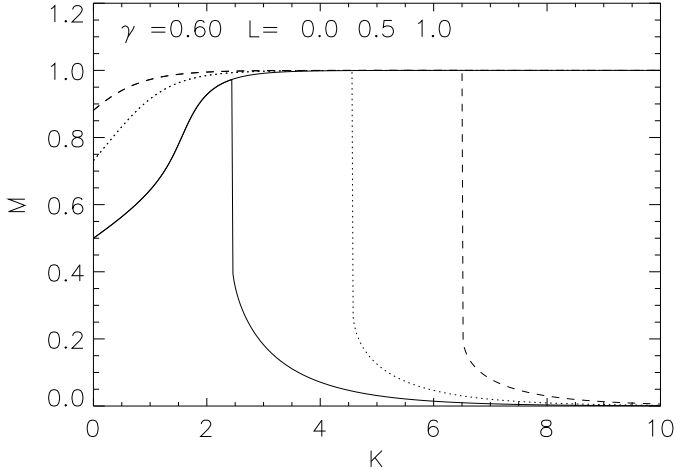}}}\fi
\quad 
\caption{The inflation parameter $M$ from eqn.\ (\ref{mtree})
         as function of $K$ for $\gamma=0.6$ and $z=3$ and various 
         $L$ values.  There exists a critical $K(L)$ where the 
         system changes from the OSP into the TSP 
         with one value $M_1\sim 1$ and one value with $M_0\sim 0$.}
\label{magnetization_b}
\end{figure}
\noindent
Choosing a value $\gamma=0.6 > \gamma_c$ we show in 
Fig.\ \ref{magnetization_b} the inflation parameter $M$
with $z=3$ as function of $K$ for various $L$ values. Above
$K_0$ with $L=L_-(K_0)>0$ the system is always in the TSP with 
$M$ values near 0 or 1 corresponding to ratios of money
values $I_n/I_{n-1}=1$ or $r_0$. Even in the deterministic limit
$K \rightarrow \infty$ the inflationary solution cannot be avoided.
The boundaries of the TSP phase in the $((K,L/K)$ plane are
are shown in Fig.\  \ref{phase_a} ($\gamma=0.25<\gamma_c$) and
Fig.\  \ref{phase_b} ($\gamma=0.6>\gamma_c$) for $z=3$.
In Fig.\  \ref{phase_a} the regions where $M$ is smaller than 0.5 (0.1)
are indicated by the dotted (dashed) line, which occur outside of
the TSP region. Therefore small values of $M$ are guaranteed
in the limit of large $K$. In contrast for $\gamma >\gamma_c$
the region of small $M$ lies entirely in the TSP region, as seen
from Fig.\  \ref{phase_b}.
\noindent The OSP can be obtained only for
negative $L$ which implies $r_0<1$ which is against the 
agents interest in the selling mode. On the other side for
$\gamma >\gamma_c$ small values of $M$ can be obtained already 
at moderate $K$.
\noindent
The phase transition crossing the bounds $L_{\pm}$
from OSP to TSP will be in general a first
order transition, since $M$ can change discontinuously by
$\Delta M$. If one approaches the end points of TSP near $K_c$
given by (\ref{Kcrit}), the discontinuity vanishes with a
power law according
\begin{eqnarray}
\Delta M \sim \left( K-K_c \right)^\frac{1}{2}
\end{eqnarray}
indicating a second order phase transition of the mean field class.
With increasing number $z$ of neighboring agents the boundaries of the
TSP degenerate into straight lines $L_+=K\gamma ^2$ and $L_-=K\gamma
(\gamma -2)$ implying presence of only the TSP for $L/K < \gamma^2$.
The same method can be applied to the buying mode, where agent $N$
buys goods via the tree from agents at $n=0$. One obtains a similar
recursion formula as (\ref{wrecursion}) with value $L',K'$ and
$\gamma '$ obtained by the replacement
\begin{eqnarray}
L'=-L\;,\; K'=\gamma^2K\;\mbox{and}\;\gamma'=1/\gamma \; .
\label{replacement}
\end{eqnarray}
This leads to qualitatively similar phase transitions.

The money value regulating authorities can achieve a stable 
economy with an 
inflation parameter $M=0$ for given agent parameters $L$ and
$\gamma$ by the choice of large $K$. The success depends on the
value of the elasticity ratio $\gamma=\beta/\alpha$.
Very different utilities $\tilde{c}(q)$ and $d(q)$
lead to $\gamma \ll 1$. In this case the system remains in the OSP
and the desired result is obtained for large $K$. For similar 
utilities $\tilde{c}(q)$ and $d(q)$ we expect $\gamma \sim 1$ and TSP
occurs with $M_0 \sim 0$ and $M_1 \sim 1$. Which solution is obtained
depends on the boundary values of the agents at $n=0$. Since their
utility function (\ref{utilityb3}) increases with $r_0$ they
prefer a value $r_0>1$ leading to increasing money values
from $n=0$ to $n=N$ (deflation). Additional measures as indirect
taxes are required to persuade the $n=0$ agents to choose the
solution $M_0$. Alternatively one can close the selling tree
by a second tree in the buying mode, where the agents at $n=0$ sell
their goods (f.ex.\ labor) through a tree to the top agents. 
In general, such a mechanism should exist in order to recycle the money
flow from $n=0$ to $n=N$ in the selling mode. In this case an
inflationary value $r_0<1$ is preferred. Combining both trees
indeed allows the intermediate state $r_0=1$ to be reached, 
as desired in a stable economy.  

\section{Conclusions}
\label{C}
In this article we considered a trading model of agents on the 
hierarchical network of a Cayley tree, treating money values as 
dynamical variables. The claim of ref. \cite{Bak} that constant
money values should result independently of geometry and utility functions
of the agents does not appear to be entirely true. Even in the case of a
linear chain, imposing money conservation at each agent we find
constant $I$, however, different in the selling and buying mode
leading to the ``peanuts effect''. Only within the periodic boundary
conditions of \cite{Bak} these constants are the same. 

When agents are allowed to choose between neighbors, as for 
$z>2$, additional dynamical phenomena may occur, dependent on 
whether agents cooperate or not. We include this as an 
optimization problem between nearest neighbors and next to 
nearest neighbors which, moving the model to the dual lattice 
(the cactus in $z=3$) still can be described in terms of nearest
neighbor interactions (now between links).  
An elegant simplification of this model in terms of an 
Ising model allows to include noise and to explicitly solve the model.   
The phases of this Ising version of the model correspond to different 
dynamical regimes of the economy. The main result is the existence of 
a two state phase above a critical money conservation parameter 
$K_c = \frac{1}{\gamma} \; \ln \frac{z}{z-2}$ 
with critical curves separating the one state phase from the two state 
phase. In the two state phase one observes a first order phase transition 
between an inflationary phase and a phase with stable money value. 
For elasticity parameters $\gamma < \gamma_c = 1-\sqrt{\frac{z-2}{z-1}}$ 
the system can remain in a one state phase with stable money value  
at $K\rightarrow\infty$, whereas for $\gamma>\gamma_c$ the two state 
phase cannot be avoided. Since $\gamma$ here is equal to the ratio 
of the exponents of the utility functions, the details of the equilibrium 
properties depend on the latter. In particular, they also depend on the 
geometry as the two state phase does not occur in the linear model. 

These findings are obtained by approximating the ratios of money 
values $I_{n+1}/I_n$ by discrete Ising variables with only two values
and, less important, using power laws for the utilities. 
The main motivation for these approximations is the possibility 
to carry out most calculations analytically. The assumption of 
power laws seems to be not too restrictive. The first assumption
of two valued variables can be relaxed by using a larger number of 
$q$ different values for $I_{n+1}/I_n$. The resulting $q$-state Potts 
model with only nearest neighbor interactions, for $K>0$ has a similar
phase structure as the Ising model \cite{Wagner}. The chaotic
behavior observed in this model \cite{Wagner2} can, not entirely 
surprising, occur only for negative $K$ where the authorities aim for
inflation. Therefore the Ising models with ferromagnetic coupling ($K>0$)
should be representative for the general case.

\end{document}